\documentclass[journal]{IEEEtran}

\ifCLASSINFOpdf
\else
\fi

\usepackage{ bbold }
\usepackage{cite}
\usepackage{xcolor}
\usepackage{graphicx}
\usepackage{amsmath}
\usepackage{amsfonts}
\usepackage{physics}
\usepackage{algorithm}

\usepackage[noend]{algpseudocode}
\usepackage{hhline}
\usepackage{ upgreek }
\usepackage{comment}
\usepackage[hidelinks]{hyperref}
\usepackage{enumitem}
\usepackage{tabularx}
\usepackage{booktabs}
\newcolumntype{C}{>{\centering\arraybackslash}X} 
\makeatletter
\def\BState{\State\hskip-\ALG@thistlm}
\makeatother

\makeatletter
\newsavebox\myboxA
\newsavebox\myboxB
\newlength\mylenA

\newcommand*\xoverline[2][0.75]{%
    \sbox{\myboxA}{$\m@th#2$}%
    \setbox\myboxB\null
    \ht\myboxB=\ht\myboxA%
    \dp\myboxB=\dp\myboxA%
    \wd\myboxB=#1\wd\myboxA
    \sbox\myboxB{$\m@th\overline{\copy\myboxB}$}
    \setlength\mylenA{\the\wd\myboxA}
    \addtolength\mylenA{-\the\wd\myboxB}%
    \ifdim\wd\myboxB<\wd\myboxA%
       \rlap{\hskip 0.5\mylenA\usebox\myboxB}{\usebox\myboxA}%
    \else
        \hskip -0.5\mylenA\rlap{\usebox\myboxA}{\hskip 0.5\mylenA\usebox\myboxB}%
    \fi}
\makeatother

\begin{document}

\title{Pixel-weighted Multi-pose Fusion for Metal Artifact Reduction in X-ray Computed Tomography}

\author{Diyu Yang$^{1}$, 
        Craig A. J. Kemp$^{2}$,
        Soumendu Majee$^{3}\textsuperscript{\textsection}$,
        Gregery T. Buzzard$^{1}$, 
        and~Charles A. Bouman$^{1}$\\
$^{1}$Purdue University-Main Campus, West Lafayette, IN 47907.\\
$^{2}$Eli Lilly and Company, Indianapolis, IN 46225.\\
$^{3}$Samsung Research America, Mountain View, CA 94043.
\thanks{Diyu Yang was supported by Eli Lilly and Company. Gregery T. Buzzard was supported by NSF grant CCF-1763896. Charles A. Bouman was supported by the Showalter Trust. }\\}

\maketitle
\begingroup\renewcommand\thefootnote{\textsection}
\footnotetext{This work was done while Soumendu Majee was employed at Purdue University.}

\begin{abstract}

X-ray computed tomography (CT) reconstructs the internal morphology of a three dimensional object from a collection of projection images, most commonly using a single rotation axis. 
However, for objects containing dense materials like metal, 
the use of a single rotation axis may leave some regions of the object obscured by the metal, even though projections from other rotation axes (or poses) might contain complementary information that would better resolve these obscured regions.  

In this paper, we propose pixel-weighted Multi-pose Fusion to reduce metal artifacts by fusing the information from complementary measurement poses into a single reconstruction.
Our method uses Multi-Agent Consensus Equilibrium (MACE), an extension of Plug-and-Play, as a framework for integrating projection data from different poses. 
A primary novelty of the proposed method is that the output of different MACE agents are fused in a pixel-weighted manner to minimize the effects of metal throughout the reconstruction.
Using real CT data on an object with and without metal inserts, we demonstrate that the proposed pixel-weighted Multi-pose Fusion method significantly reduces metal artifacts relative to single-pose reconstructions. 

\end{abstract}


\IEEEpeerreviewmaketitle

\section{Introduction}
\label{sec:intro}
X-ray computed tomography (CT) imaging is widely used in industrial \cite{sun2012overview} and medical \cite{hiriyannaiah1997x} applications for non-destructive visualization of internal sample morphology.
X-ray CT uses a series of projection images from various angles  to reconstruct a 3D array of attenuation coefficients that describe the sample \cite{withers2021x}.
Traditional CT reconstruction methods, such as Filtered Back Projection (FBP), use projection images acquired around a single rotation axis. 
However, for some objects, projection images from different rotation axes may contain complementary information. 
In such cases, more accurate reconstructions could in principle be achieved by collecting projection images from multiple rotation axes (or poses), and then performing a joint or multi-pose reconstruction. 

Multi-pose reconstruction is motivated by samples with dense materials such as metal, which produce substantial, spectrally-dependent attenuation in the projection data. Standard reconstructions from such data exhibit bright or dark streaks, commonly known as metal artifacts, radiating out from the metal object \cite{MARsummary}.
These metal artifacts are strongly influenced by the measurement pose of the object \cite{MARsummary, Brown1999, Lewis2010, luckow2011tilting}, and specific regions of the object may be corrupted in one pose but artifact-free in another. 
Therefore, multi-pose reconstruction holds the potential to mitigate metal artifacts by effectively taking the best quality information from each pose.

One approach to multi-pose reconstruction is to transform separate reconstructions to a common pose and fuse them using a post-processing technique such as averaging \cite{ballhausen2014post}, pixel-wise minimum/maximum \cite{guhathakurta2015reducing}, or weighted sum \cite{kobayashi2017multi,herl2018metal}. 
However, this approach operates in the image domain and hence does not fully exploit the complementary information in sinogram  measurements.
In contrast, Kostenko et al. \cite{kostenko2018registration} and Herl et al. \cite{herl2019artifact} adapt the scanner geometry of multiple poses to a common pose to produce a joint reconstruction from sinogram data. 
However, this method is complex to implement, since it requires reprogramming the system matrix and integrating pose information into the reconstruction software. 

Metal artifacts can also be reduced by the use of Model-based Iterative Reconstruction (MBIR) \cite{MBIR, thibault2007three, zhang2013model, jin2015model, kisner2012model, jordanweight, balke2018separable},
which leverages the regularity within reconstructions to compensate for noisy or limited measurements. 
More recently, MBIR has evolved to include 
Plug-and-Play (PnP) \cite{PnPPrior, PnPPriorSuhas}, which uses a denoiser to model the prior distribution of an image, and Multi-Agent Consensus Equilibrium (MACE), which allows multiple agents to represent various objectives in an inverse problem \cite{CE, DistributedMACE, MACE4D, FCI2022, mutli-pose}.  
One notable advantage of MACE is its formulation as an equilibrium problem, making it applicable even in cases where there is no well-defined cost function to minimize.
In this context, Yang et al. introduced a Multi-pose Fusion approach to perform a joint tomographic reconstruction using multiple poses of a single object \cite{mutli-pose}.

\begin{figure}[t]
\centering
   \includegraphics[width=\linewidth]{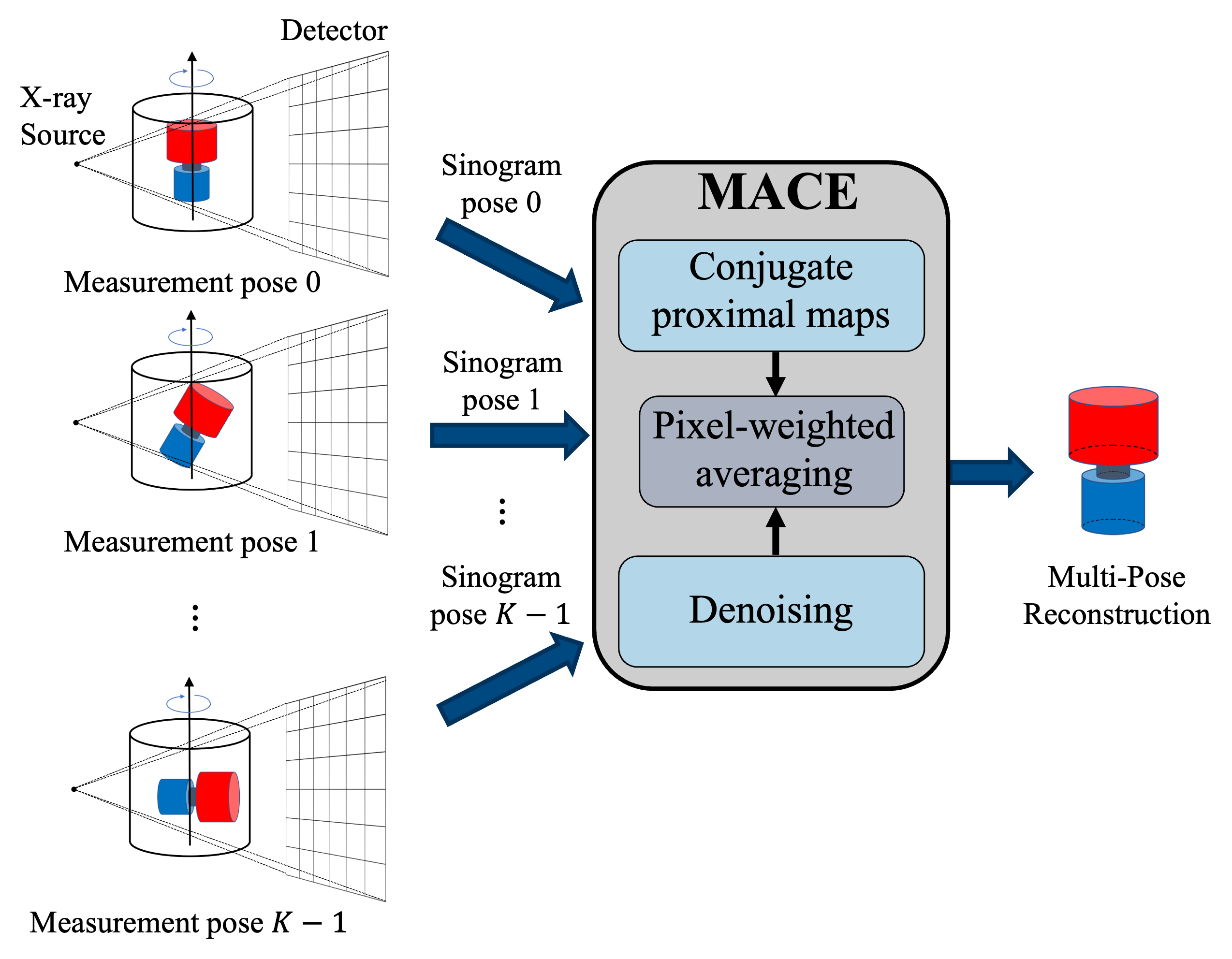}
   \caption{Multi-pose Fusion Overview: Multiple sets of CT measurements are acquired from different poses of the same object. The measurement data from different poses are then fused in a MACE framework to form a joint reconstruction.}
   \label{fig:MPF-overview}
\end{figure}


In this paper, we build on the prior work of \cite{mutli-pose}, and propose a pixel-weighted Multi-pose Fusion algorithm, which reduces metal artifacts by integrating CT measurements from multiple poses of a single object (see Figure \ref{fig:MPF-overview}).
As illustrated in Fig. \ref{fig:MPF-overview}, the proposed algorithm uses MACE to integrate information from the various poses, with each pose being represented by a single agent. 
The final reconstruction is then computed as the fixed point of a set of equilibrium equations.
This allows for a straightforward, modular implementation using standard CT reconstruction software.
A primary novelty of the proposed method is that the output of different MACE agents are fused in a pixel-weighted manner to minimize the effects of metal throughout the reconstruction.
We present experimental results for the problem of non-destructive evaluation (NDE) using measured CT data, and demonstrate that pixel-weighted Multi-pose Fusion method is effective in reducing metal artifacts and improving image quality.

\section{Problem Formulation}\label{sec:problem_formulation}

In multi-pose CT imaging, multiple sets of CT scans are taken using different poses of the object, as illustrated in Fig. \ref{fig:MPF-overview}. 
Notice that the imaging geometry is the same for each pose.
However, in practice some poses provide more useful information, particularly when the object contains dense or even opaque components that may obscure portions of the object.
The objective of multi-pose CT reconstruction is then to perform a joint MBIR reconstruction from scans acquired from multiple poses of the object. 

Let $y_k \in \mathbb{R}^{M_k}$ be the sinogram measurements for the $k^{th}$ pose, where $k\in\left\{0,...,K-1\right\}$.
Then our goal is to recover $x\in\mathbb{R}^{N}$, the image vector containing attenuation coefficients in the reconstruction coordinate system.

For each pose $k$, we also define a transformation function $x_k = T_k x$ where $x$ is the object represented in the common reconstruction coordinate system and $x_k$ is the object represented in the $k^{th}$ pose.
So intuitively, $T_k$ transforms the raster sampled object from the common reconstruction coordinate system to the posed coordinate system.
In practice, $T_k$ typically implements a rigid body transformation \cite{rigidtransform}, so it requires that the discretized function be resampled on the transformed sampling grid. 
This process requires some form of typically spline-based interpolation algorithm \cite{de1978practical}.
We will also require an approximate inverse transformation $T_k^{-1}$. 
Since both transforms require interpolation and resampling, we note that they will not in general be exact inverses of each other.

Using this notation, the forward model for each pose has the form 
\begin{equation}
    y_k = A_k T_k x + w_k \ ,
\end{equation}
where $A_k \in \mathbb{R}^{N\times M_k}$ is the scanner system matrix and $w_k\sim N(0, \alpha \Lambda_k^{-1})$ is independent additive noise, each for the $k^{th}$ pose.

The joint MBIR reconstruction for the multi-pose problem is then given by
\begin{equation}
    \label{eq:inverse_problem}
    x^* = \arg\min_{x}\left\{ \sum_{k=0}^{K-1} f_k(x)+h(x)\right\}
\end{equation}
with data fidelity terms given by $f_k (x) = -\log p(y_k |x) + \mbox{const}$ where
\begin{equation}
   \label{eq:pose_cost}
   f_k (x) = \frac{1}{2 } \lVert  y_k-A_kT_k x  \rVert_{\Lambda_k}^2 \ ,
\end{equation}
and a prior term given by $h(x)=-\log p(x)$ that imposes regularity.

Notice that direct implementation of \eqref{eq:inverse_problem} is difficult since it requires that software be written to minimize a sum of complex tomographic reconstruction terms each with a different transformation $T_k$.

Alternatively, one can compute the MBIR reconstruction by using consensus ADMM \cite{BoydADMM} to minimize the sum of $K+1$ terms consisting of $h$ and the $K$ terms in \eqref{eq:pose_cost}.
However, this approach has two serious disadvantages.
First, the proximal map terms required for each pose will be very computationally expensive to compute. 
Second, we can improve reconstruction quality by replacing the prior term $h(x)$ with a PnP denoiser.

\section{MACE formulation of Pixel-weighted Multi-pose Fusion}
\label{sec:mace_agents}

In this section, we introduce the MACE framework for solving the multi-pose reconstruction problem \cite{CE,FCI2022}.

\subsection{Agent formulation}
We start by describing the agents in our approach.
Let $x^\prime = F_k (x)$ be an agent for the $k^{th}$ pose.
Intuitively, this agent should take a reconstruction $x$ and return a reconstruction $x^\prime$ that better fits the measurements $y_k$ associated with the data from pose $k$.
One approach would be to use \eqref{eq:pose_cost} directly in a proximal map.  
However, this would be computationally expensive and difficult to compute since it requires that the transformation $T_k$ be integrated into the reconstruction software.

Alternatively, we propose to use a \emph{Conjugate Proximal Map} given by
\begin{equation}
    \label{eq:CGPM_final}
    F_k (v) = T_k^{-1} F (T_k v ; y_k ) \ ,
\end{equation}
where $F(v; y)$ is the standard proximal map in reconstruction coordinates given by
\begin{equation}
    \label{eq:standard_prox}
    F(v; y ) = \arg\min_{x} \left\{ \frac{1}{2} \lVert y-A_k x \rVert_{\Lambda_k}^2 + \frac{1}{2\sigma^2}\lVert x - v\rVert^2\right\}.
\end{equation}
Notice that the conjugate proximal map of \eqref{eq:CGPM_final} can be computed easily since it requires only the computation of the standard proximal map of \eqref{eq:standard_prox} in the standard coordinates and pre- and post-composition with the spline-based maps $T_k$ and $T_k^{-1}$.
In fact, software for computing the proximal map in \eqref{eq:standard_prox} is openly available \cite{svmbir, mbircone}.

The conjugate proximal map is exactly equivalent to the conventional proximal map based on \eqref{eq:pose_cost} when the rigid body transformations $T_k$ and $T_k^{-1}$ are exact inverses of each other (see \hyperref[appendix:conj_prox_map]{Appendix A}). 
In practice, we note that $T_k$ and $T_k^{-1}$ will generally not be exact inverses due to the nonlinear interpolation, hence the solution to the conjugate proximal map is an approximation to the conventional proximal map.

For the prior model, we will use a BM4D denoiser \cite{maggioni2012nonlocal}. We denote this agent by $F_K$.

The MACE agents are concatenated together to form a single operator ${\bf F}: \mathbb{R}^{(K+1)N} \rightarrow \mathbb{R}^{(K+1)N}$, defined as:
\begin{equation}
\label{eq:stacked_agents}
{\bf F} ( {\bf x})= \left[ F_0(x_0), \cdots ,F_{K}(x_{K})\right],
\end{equation}
where ${\bf x}=\left[ x_0,...,x_K\right]$ denotes the full MACE state.

\subsection{Pixel-weighted MACE}
The MACE state vector ${\bf x}$ contains multiple, potentially inconsistent reconstructions $x_0,...,x_K$. 
In order to produce a single coherent MACE reconstruction, we define a pixel-weighted averaging operator 
\begin{equation}
\label{eq:weighted_consensus_operator}
{\bf G_M} ( {\bf x})= \left[ \bar{x}_{\bf M}({\bf x}) , \cdots , \bar{x}_{\bf M}({\bf x}) \right] \ ,
\end{equation}
where $\bar{x}_{\bf M}({\bf x})$ is a pixel-weighted average of the input vector components given by
\begin{equation}
\label{eq:weighted_avg}
\bar{x}_{\bf M}({\bf x})=\frac{1}{1+\beta} \sum_{k=0}^{K-1}M_kx_k + \frac{\beta}{1+\beta}x_{K}.
\end{equation}
Here, $\beta>0$ controls the amount of regularization relative to the data-fitting agents,  and $M_k \in \mathbb{R}^{N \times N}$ is a diagonal weight matrix specific to each data-fitting agent satisfying the property
$$
\begin{cases}
    \left[ M_k\right]_{ii}  \geq 0 \\
    \sum_{k=0}^{K-1} M_k = I \\
\end{cases} \ .
$$
Intuitively, ${\bf G}_{\bf M}({\bf x})$ computes a weighted average of the components in ${\bf x}$ according to the weight matrices $M_0, ..., M_{K-1}$ as well as the regularization parameter $\beta$, and then returns a state vector formed by replicating this average $K+1$ times.

Notice that $M_k$ provides a mechanism to weight each pixel in each pose separately, which could be valuable in the metal artifact scenario, where certain components from specific poses are corrupted.
The design of this weight matrix is discussed in Section \ref{sec:pixel_weighted_averaging}.

Using this notation, the MACE equilibrium equation is
\begin{equation}
\label{eq:MACE}
{\bf F} ( {\bf x^*})= {\bf G_M} ( {\bf x^*}) \, 
\end{equation}
where ${\bf x^*}$ solves the equation, and the final reconstruction is then given by $x^*=\bar{x}_{\bf M}({\bf x^*})$.
This equation enforces that all agents have the same output value (consensus) and that the vectors $\delta^*_k = x^*_k - F_k(x^*_k)$ satisfy $\bar{x}_{\bf M}({\bf \delta}) = 0$ (equilibrium) \cite{CE}.  

\subsection{Computing the MACE Solution}

It can be shown that the solution to (\ref{eq:MACE}) is also the fixed point of the operator ${\bf T}=(2{\bf G_M}-I)(2{\bf F}-I)$ (see \hyperref[appendix:MACE_fixed_point]{Appendix B}). 
One popular method of finding such a fixed point is Mann iteration
\begin{equation}
    {\bf w} \leftarrow(1-\rho) {\bf w}+\rho {\bf T} {\bf w},
\end{equation}
where $\rho\in (0,1)$ controls the convergence speed.

\begin{algorithm}
\caption{Pixel-weighted MACE algorithm}\label{MACE_algorithm}
\textbf{Input}: Initial Reconstruction: $x^{(0)} \in \mathbb{R}^N$ \\
\textbf{Input}: Weight matrices $M_0,...,M_{K} \in\mathbb{R}^{N\times N}$\\
\textbf{Output}: Final Reconstruction: $x^* \in \mathbb{R}^N$
\begin{algorithmic}[1]
\State ${\bf w} \leftarrow[x^{(0)},...,x^{(0)}]$
\While {not converged}
\State ${\bf x} \leftarrow {\bf F} ( {\bf w} )$
\State ${\bf z} \leftarrow {\bf G_M} ( 2{\bf x} - {\bf w})$
\State ${\bf w} \leftarrow {\bf w} + 2\rho({\bf z}-{\bf x})$
\EndWhile
\State \Return $x^*\leftarrow \bar{x}_{\bf M}({\bf x})$
\end{algorithmic}
\end{algorithm}

Algorithm \ref{MACE_algorithm} shows the general method of solving MACE with Mann iterations.
The algorithm starts from an initial reconstruction $x^{(0)}$, and uses Mann iterations to find the equilibrium point between the prior and forward model terms. 
From \cite{CE}, when the agents $F_0,...,F_{K}$ are all proximal maps of associated cost functions $f_0,...,f_{K-1},h$, and all agents are equally weighted ($M_k=\frac{1}{K+1}I$ for $k\in \left\{0,...,K-1\right\}$), this equilibrium point is exactly the solution to the consensus optimization problem of (\ref{eq:inverse_problem}).

\section{Pixel-weighted Averaging for Metal Artifact Reduction}
\label{sec:pixel_weighted_averaging}

We leverage the weight matrix $M_k$ in \eqref{eq:weighted_avg}, which provides a mechanism of applying a separate weight to each pixel in each pose, to propose a pixel-weighted averaging algorithm for metal artifact reduction.

Given an initial reconstruction $x^{(0)}$ in the reconstruction coordinates, we locate the metal and object components through the binary masks $b^{metal}, b^{object} \in \mathbb{R}^N$, where 1 indicates a metal/object pixel, and 0 elsewhere:
\begin{equation}
    \label{eq:metal_mask}
    \left[b^{metal}\right]_i=\begin{cases}
            1, & \left[x^{(0)}\right]_i>\tau_{metal}\\
            0, &  \left[x^{(0)}\right]_i \leq \tau_{metal}\\
        \end{cases}
\end{equation}

\begin{equation}
    \label{eq:obj_mask}
    \left[b^{object}\right]_i=\begin{cases}
            1, & \left[x^{(0)}\right]_i>\tau_{object}\\
            0, &  \left[x^{(0)}\right]_i \leq \tau_{object}\\
        \end{cases}
\end{equation}
where $\tau_{metal},\tau_{object} \in \mathbb{R}$ are threshold values to identify the metal and object pixels.
We then transform the image masks from the reconstruction pose to each of the measurement poses $k \in \left\{0,..., K-1 \right\}$:
\begin{equation}
    \label{eq:metal_mask_posed}
    b_k^{metal} = T_k b^{metal}
\end{equation}
\begin{equation}
    \label{eq:obj_mask_posed}
    b_k^{object} = T_k b^{object}
\end{equation}

For each pose $k$, we compute a distortion image $D_k \in \mathbb{R}^N$, which estimates the level of distortion at each pixel in the associated measurement pose:
\begin{equation}
\label{eq:distortion_matrix}
    \left[D_k\right]_i = \frac{\left[A_k^tA_kb_k^{metal}\right]_i}{\left[A_k^tA_kb_k^{object}\right]_i+\epsilon},
\end{equation}
where small $\epsilon>0$ prevents division by zero.
Here, $D_k$ uses both the location of the metal components ($b_k^{metal}$) and the scanner geometry ($A_k$) to predict the pixel-wise level of distortion.
Intuitively, a larger entry in $D_k$ indicates that the associated projections traverse a longer distance through metal, which could lead to more corruptions in the reconstructed pixel.

The weight matrices $M_0, ..., M_{K-1}$ are then calculated using a softmax function across all distortion images:
\begin{equation}
\label{eq:weight_matrix}
    \left[M_k\right]_{ii} = \frac{\exp{-\alpha\left[T_k^{-1}D_k\right]_i}}{\sum_{m=0}^{K-1}\exp{-\alpha \left[T_m^{-1}D_m\right]_i}},
\end{equation}
where $T_k^{-1}$ transforms the distortion image from the measurement pose to the common reconstruction pose, and $\alpha \geq 0$ controls the range of the weight matrices.

Subsequently, pixel-weighted Multi-pose Fusion (Algorithm \ref{MACE_algorithm}) uses these weight matrices to produce a joint reconstruction that selectively fits the informative measurements from different poses.

Notice that the proposed algorithm can also work as a post-processing method to directly integrate the images from different poses.
We call this method the pixel-weighted post-processing.
The pseudo-code for this method is depicted in Algorithm \ref{algo:pixel-weighted-average}.
The algorithm takes standard CT reconstructions $x_0,...,x_{K-1}$ as inputs, each from a distinct pose. For each pose, the distortion image $D_k$ and subsequently the weight matrix $M_k$ are computed.
The post-processed image is formed by taking the pixel-weighted average among the input reconstructions $x_0,...,x_{K-1}$. 

\begin{algorithm}
\caption{Pixel-weighted Post-processing for Metal Artifact Reduction}\label{algo:pixel-weighted-average}
\textbf{Input}: Standard reconstructions: $x_0, ..., x_{K-1}\in\mathbb{R}^{N}$ \\
\textbf{Input}: Transformation mappings: $T_0, ..., T_{K-1}$ \\
\textbf{Output}: Pixel-weighted average: $x^* \in \mathbb{R}^{N}$
\begin{algorithmic}[1]
\For{$k\leftarrow 0$ to $K-1$}
\State Compute $b_{k}^{metal}$, $b_{k}^{object}$  \Comment{image masks Eq. \eqref{eq:metal_mask_posed}, \eqref{eq:obj_mask_posed}}
\State Compute $D_k$ \Comment{distortion image, Eq. \eqref{eq:distortion_matrix}}
\State Compute $M_{k}$ \Comment{weight matrix, Eq. \eqref{eq:weight_matrix}}
\EndFor
\State \Return $x^{*}=\sum_{k=0}^{K-1}M_kT_k^{-1}x_k$ 
\end{algorithmic}
\end{algorithm}

\section{Experimental Results}
\label{sec:results}
We evaluate the effectiveness of pixel-weighted Multi-pose Fusion on a real cone beam CT dataset featuring two distinct measurement poses.

\begin{figure}[t]
\centering
   \includegraphics[width=\linewidth]{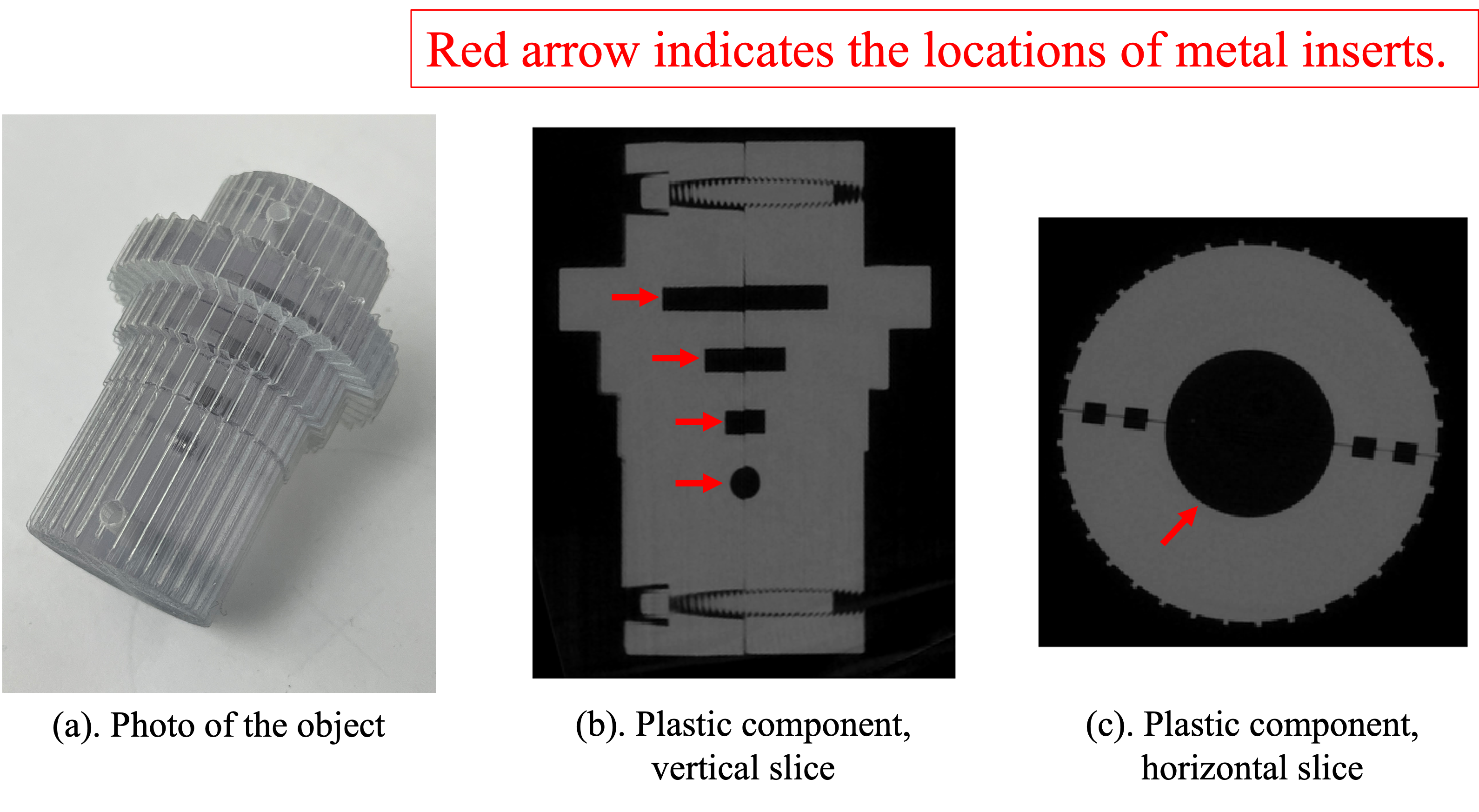}
   \caption{Demonstration of the object of interest. 
            The object is composed of a plastic component with four removable metal disks.
            (a) A photo of the object of interest.
            (b) and (c) vertical and horizontal slices without the metal components. 
            In the scans below, metal disks are inserted in the regions indicated by the red arrows. 
            }
   \label{fig:ref_image}
\end{figure}

\begin{figure}[t]
\centering
   \includegraphics[width=0.9\linewidth]{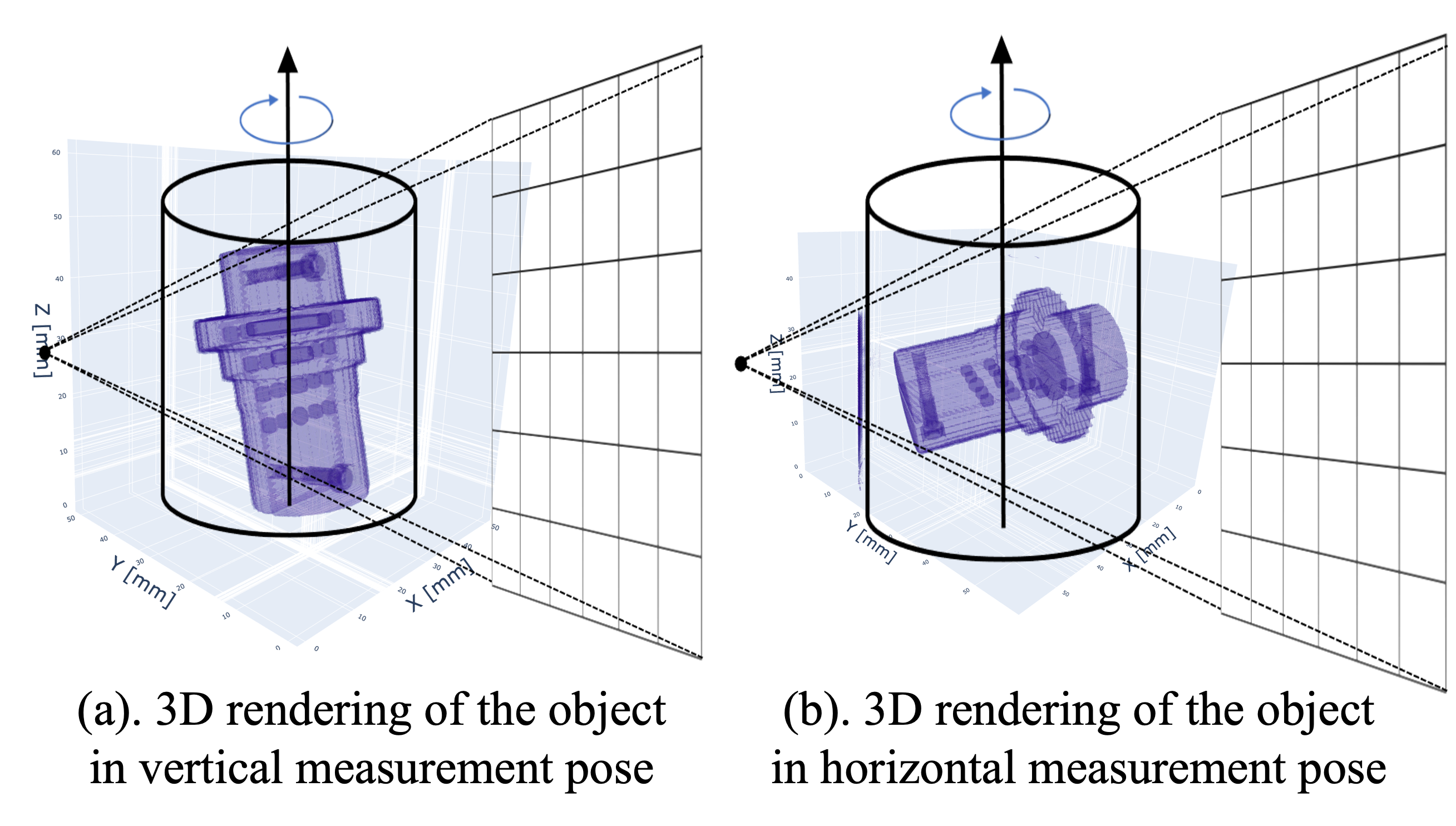}
   \caption{3D Visualization of the measurement poses. The CT scans are acquired from two measurement poses:
            (a) vertical pose. (b) horizontal pose.
            }
   \label{fig:poses_mar}
\end{figure}

\begin{table}
\caption{Experimental Specifications for Plastic Part with Metal Disk Dataset}
\label{tab:specs_mar}
\begin{tabularx}{0.45\textwidth}{@{} l *{2}{C} c @{}}
\toprule
labels  
& Vertical Pose Scan & Horizontal Pose Scan  \\ 
\midrule
Source-Detector Distance & 569.363 mm & 580.247 mm\\

Magnification & 3.87 & 3.47\\

Number of Views & 900 & 1150\\

Detector Array Size & 960$\times$768 & 768$\times$960\\

Detector Pixel Pitch & 0.254 mm & 0.254 mm\\

\bottomrule
\end{tabularx}
\end{table}

The object of interest (Fig. \ref{fig:ref_image}) is a plastic component with four removable metal disks.
The object is scanned from two different poses (Fig. \ref{fig:poses_mar}) using a North Star Imaging X50 X-ray CT system. 
The experimental specifications are detailed in Table \ref{tab:specs_mar}.

\begin{figure*}
\centering
   \includegraphics[width=\textwidth]{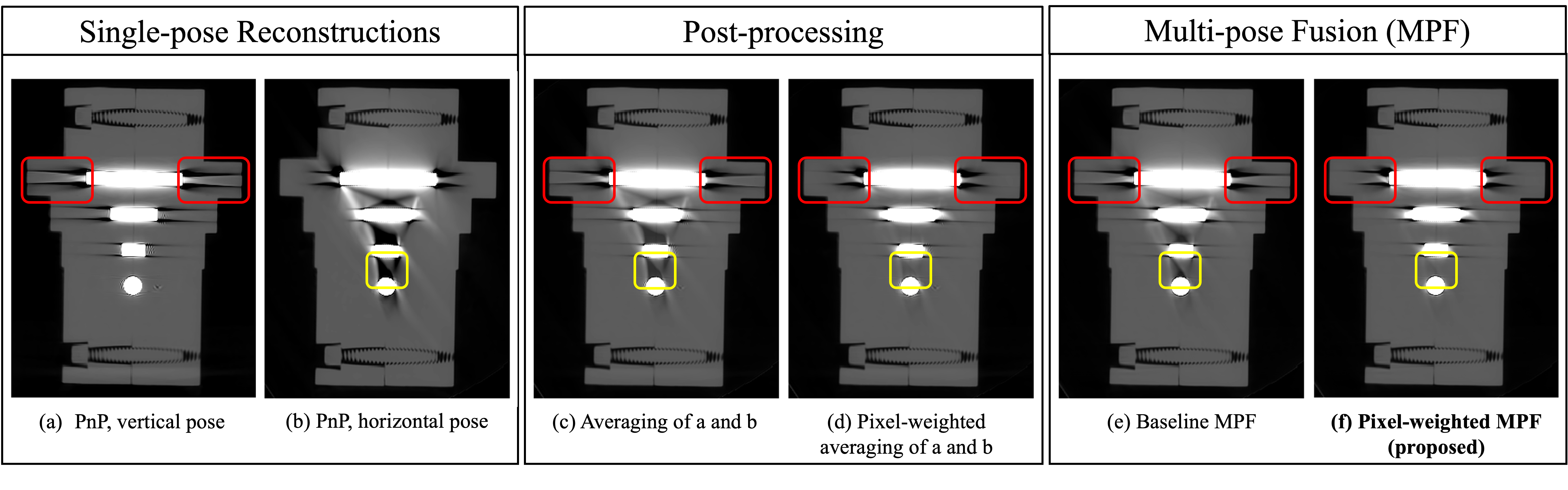}
   \caption{Comparison of different reconstruction methods and different poses.
   (a) PnP, vertical pose.
   (b) PnP, horizontal pose.
   (c) Averaging of a and b.
   (d) Pixel-weighted averaging of a and b.
   (e) Baseline Multi-pose Fusion.
   (f) Pixel-weighted Multi-pose Fusion (proposed).\\
   Pixel-weighted Multi-pose Fusion produces the best image quality with significantly reduced metal artifacts.
   }
   \label{fig:mpf_trans}
\end{figure*}

We perform a pixel-weighted Multi-pose Fusion with Algorithm \ref{MACE_algorithm}, and compare the results with various reconstruction and post-processing methods listed below:
\begin{enumerate}[label=(\alph*)]
    \item PnP, vertical pose: PnP recon from vertical pose.
    \item PnP, horizontal pose: PnP recon from horizontal pose.
    \item Averaging: Averaging of the PnP results.
    \item Pixel-weighted averaging: Pixel-weighted averaging of the PnP results.
    \item Baseline MPF: Baseline Multi-pose Fusion. The agents are equally weighted.
    \item {\bf Pixel-weighted MPF (proposed)}: Pixel-weighted Multi-pose Fusion. A separate weight is assigned to each pixel in each agent.
\end{enumerate}
For a fair comparison, we use the same BM4D denoiser in PnP and Multi-pose Fusion algorithms.

Fig.~\ref{fig:mpf_trans} shows the results with different methods.
The comparison of PnP results reveals varying metal artifact characteristics across different poses.
The artifacts mainly manifest as horizontal streaks in the vertical pose (in red boxes), while appearing as shadowy artifacts between adjacent metal disks in the horizontal pose (in yellow boxes).
Compared to simple averaging, pixel-weighted averaging further reduces metal artifacts by selectively integrating the informative components from different poses.
Pixel-weighted Multi-pose Fusion further reduces metal artifacts by incorporating this pixel-weighted averaging mechanism into MACE framework, which produces the best image quality with significantly reduced metal artifacts.

\section{Conclusion}
In this paper, we have introduced a novel CT reconstruction method called pixel-weighted Multi-pose Fusion. 
Our method uses Multi-Agent Consensus Equilibrium (MACE), an extension of Plug-and-Play, as a framework for integrating projection data from different poses. 
A primary novelty of the proposed method is that the output of different MACE agents are fused in a pixel-weighted manner to minimize the effects of metal throughout the reconstruction.
Our experiment demonstrated that pixel-weighted Multi-pose Fusion delivers a significant reduction in metal artifacts compared to single-pose reconstruction and post-processing methods.

\section*{Appendix}
\label{sec:appendix}

\subsection{Relationship between Conventional Proximal Map and Conjugate Proximal Map}
\label{appendix:conj_prox_map}
The conjugate proximal map for the $k^{th}$ pose is given by
\begin{equation}
    \label{eq:conjugate_PM_expanded}
    F_k (v) = T_k^{-1}\arg\min_{x}\left\{ \frac{1}{2}\lVert y_k-Ax \rVert_{\Lambda_k}^2 + \frac{1}{2\sigma^2}\lVert x-T_kv\rVert^2\right\}.
\end{equation}
We show that the conjugate proximal map \eqref{eq:conjugate_PM_expanded} is exactly equivalent to the conventional proximal map based on \eqref{eq:pose_cost} when the transformation mappings $T_k, T_k^{-1}$ satisfy the following conditions for all $x\in \mathbb{R}^{N}$:
\begin{enumerate}[label={(\alph*)}]
    \item $x=T_k^{-1}T_kx$ (inverse condition).   \label{exact_inverse}
    \item $\lVert x \rVert = \lVert T_k x\rVert$ (rigid body transformation) \label{rigid_transform}
\end{enumerate}

With a change of variable $x=T_kx_k$, we may rewrite the conjugate proximal map \eqref{eq:conjugate_PM_expanded} as
\small
\begin{align}
    F_k (v) &= T_k^{-1}\arg\min_{T_kx_k}\left\{ \frac{1}{2}\lVert y_k-AT_kx_k \rVert_{\Lambda_k}^2 + \frac{1}{2\sigma^2}\lVert T_kx_k-T_kv\rVert^2\right\} \\
    \label{CGPM_derivation_2}
    &= \arg\min_{x_k}\left\{ \frac{1}{2}\lVert y_k-AT_kx_k \rVert_{\Lambda_k}^2 + \frac{1}{2\sigma^2}\lVert T_k(x_k-v)\rVert^2\right\} \\
    \label{CGPM_derivation_3}
    &= \arg\min_{x_k}\left\{ \frac{1}{2}\lVert y_k-AT_kx_k \rVert_{\Lambda_k}^2 + \frac{1}{2\sigma^2}\lVert x_k-v\rVert^2\right\},
\end{align}
\normalsize
where \eqref{CGPM_derivation_2} requires the inverse condition \ref{exact_inverse}, and \eqref{CGPM_derivation_3} requires the rigid body transformation condition \ref{rigid_transform}.

Notice that after substituting the dummy variable $x_k$ with $x$ in \eqref{CGPM_derivation_3}, the conjugate proximal map is exactly equivalent to the conventional proximal map:
\begin{equation}
    \tilde{F}_k (v) = \arg\min_{x}\left\{ \frac{1}{2} \lVert y-A_k T_k x \rVert_{\Lambda_k}^2 + \frac{1}{2\sigma^2}\lVert x-v\rVert^2\right\}
\end{equation}

\subsection{Reformulating MACE as a Fixed-point Problem}
\label{appendix:MACE_fixed_point}
We follow \cite{CE} and show that MACE equation \eqref{eq:MACE} can be reformulated as a fixed point problem.

For notation simplicity, we rewrite the weight matrices as
$$
M_k^{\prime} = 
\begin{cases}
    \frac{1}{1+\beta}M_k , &0 \leq k \leq K-1 \\
    \frac{\beta}{1+\beta}I, &k=K
\end{cases} \ .
$$
With this notation, the weighted averaging operator \eqref{eq:weighted_avg} can be rewritten as: 
\begin{equation}
    \bar{x}_{\bf M}({\bf x})=\sum_{k=0}^{K}M_k^{\prime} x_k.
\end{equation}

By definitions of $\bar{x}_{\bf M}$ and ${\bf G}_{\bf M}$, we have
\begin{align}
    \bar{x}_{\bf M}({\bf G}_{\bf M}({\bf x})) &= \sum_{k=0}^{K} M_k^{\prime}\left[ {\bf G}_{\bf M}({\bf x}) \right]_{k}\\
    \label{fixed_point_derivation_1}
    &= \sum_{k=0}^{K} M_k^{\prime} \bar{x}_{\bf M}({\bf x}) \\
    \label{fixed_point_derivation_2}
    &= \bar{x}_{\bf M}({\bf x}),
\end{align}
where \eqref{fixed_point_derivation_2} holds because $\sum_{k=0}^K M_k^{\prime} = I$.
This gives the following property:
\begin{equation}
    {\bf G}_{\bf M}({\bf G}_{\bf M}({\bf x})) = {\bf G}_{\bf M}({\bf x})
\end{equation}
With this linear property, we may further show that ${\bf G}_{\bf M}$ satisfies the property
\begin{equation}
    (2{\bf G}_{\bf M}-{\bf I})^{-1} = (2{\bf G}_{\bf M}-{\bf I}).
\end{equation}
From this, we may reformulate the MACE equation \eqref{eq:MACE} as a fixed point problem:
\begin{equation}
    \label{fixed_point}
    (2{\bf G}_{\bf M}-{\bf I})(2{\bf F}-{\bf I})({\bf x^*}) = {\bf x^*},
\end{equation}
or ${\bf T x^*}={\bf x^*}$, where ${\bf T}=(2{\bf G}_{\bf M}-{\bf I})(2{\bf F}-{\bf I})$.


\ifCLASSOPTIONcaptionsoff
  \newpage
\fi

\bibliographystyle{IEEEtran}
\bibliography{multipose}

\end{document}